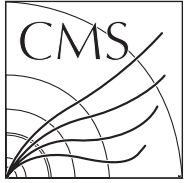

**The Compact Muon Solenoid Experiment**

# Conference Report

Mailing address: CMS CERN, CH-1211 GENEVA 23, Switzerland

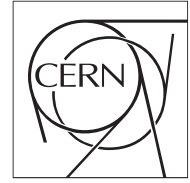

# Data Quality Monitoring of the CMS Tracker

Volker Adler for the CMS Tracker Collaboration

### Abstract

The Data Quality Monitoring (DQM) of the Compact Muon Solenoid (CMS) silicon tracking detectors (Tracker) at the Large Hadron Collider (LHC) at CERN is a software based system designed to monitor the detector and reconstruction performance, to identify problems and to certify the collected data for physics analysis. It uses the framework provided by the central CMS DQM as well as tools developed especially for the CMS Tracker DQM. This paper describes aim, framework conditions, tools and work flows of the CMS Tracker DQM and shows examples of its successful use during the recent commissioning phase of the CMS experiment.

# Data Quality Monitoring of the CMS Tracker

Volker Adler, *Vrije Universiteit Brussel* – on behalf of the CMS Tracker Collaboration

*Abstract* – The Data Quality Monitoring (DQM) of the Compact Muon Solenoid (CMS) silicon tracking detectors (Tracker) at the Large Hadron Collider (LHC) at CERN is a software based system designed to monitor the detector and reconstruction performance, to identify problems and to certify the collected data for physics analysis. It uses the framework provided by the central CMS DQM as well as tools developed especially for the CMS Tracker DQM. This paper describes aim, framework conditions, tools and work flows of the CMS Tracker DQM and shows examples of its successful use during the recent commissioning phase of the CMS experiment.

## I. Introduction

THE Data Quality Monitoring (*DQM*) system of the Compact Muon Solenoid (*CMS*) silicon tracking detectors (*Tracker*) has the two main tasks to (a) monitor the detector and reconstruction performance in order to find problems and (b) to provide data certification as basis for data selection for physics analysis. On the one hand, monitoring and problem identification have to allow for immediate feedback during data taking (*online*), so that any problem can be analyzed and solved quickly in order to record as much good-quality data as possible. It also requires a comprehensive view of the data with respect to its source in the detector and its evolution during processing. On the other hand, data certification should take into account all available information, that can make a statement about the data quality, where this information is partially only available during or after a full reconstruction of the data (*offline*). In the approach to these tasks, the CMS Tracker DQM has to take the specific design of the detector and also the framework given by the central CMS DQM into account.

## II. The CMS Tracker

The CMS Tracker consists of two sub-detectors to cope with the main tasks of a tracking system in high-energy physics: (a) a silicon pixel detector (*Pixel*) close to the interaction point to identify vertices and (b) a surrounding silicon strip detector (*SiStrip*) to measure momenta of charged particles. Both have their sensors arranged in cylindrical layers in the central region (barrel) and symmetrically in disks in both end-cap regions (forward). Fig. 1 shows the schematic layout of the CMS Tracker. A detailed description of the CMS Tracker can be found in [1].

Further segmentation of the layers and disks leads finally to the smallest monitored detector units (full detector granularity), the modules. In total, the SiStrip consists of 15,148 modules with about 9.3 million readout channels, and the Pixel has about 1,440 modules with about 66 million channels.

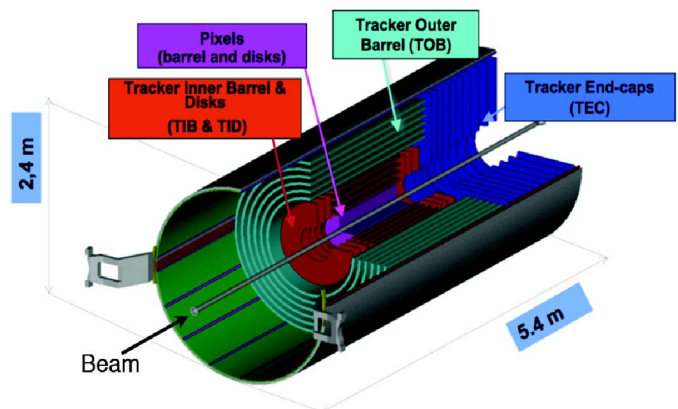

Fig. 1. Schematic layout of the CMS Tracker. The Pixel surrounds the beam pipe with three layers in the barrel and two disks per end cap These are encapsulated by the SiStrip inner detectors Tracker Inner Barrel (TIB) with four layers and Tracker Inner Disks (TID) with three disks on each side and then by the SiStrip outer detectors Tracker Outer Barrel (TOB) with six layers and Tracker End Caps (TEC) with nine disks each.

## III. The CMS DQM System

The CMS DQM system uses histograms to monitor the data quality. It provides the software infrastructure to define, fill, store and display them. The content of the histograms is computed from data by the *DQM producers*. These are software modules, which book and fill the histograms. In order to assess the quality of the data, a set of predefined quality tests can be applied to histograms by the *DQM consumers*, which are software modules accessing the histograms to perform analysis tasks including *quality tests*. These quality tests involve e.g. range checking or comparison to a reference. This procedure is applied at all levels of detector granularity and the quality test results (Boolean or float type, the latter representing a percentage) are combined to compact *quality flags*, which are used to certify the data.

The CMS DQM output is stored in ROOT files with a hierarchical folder structure. The particular structure in the sub-system folders is defined by the sub-systems (sub-detectors and basic reconstruction objects like e.g. jets or muons).

The CMS DQM system also provides web based tools for visualization of histograms and bookkeeping of DQM results. The visualization tool is the CMS DQM Graphical User Interface (GUI), which displays the histograms of all sub-systems and allows especially the central CMS DQM shifters to have a quick and comprehensive overview over the DQM status of the whole experiment. An example view of SiStrip summary histograms is shown in Fig. 2. Another tool provided by the CMS DQM, which is used in the certification process is the *Run Registry*, a database plus web client for bookkeeping of DQM certificates. Details on both tools are found in [2].

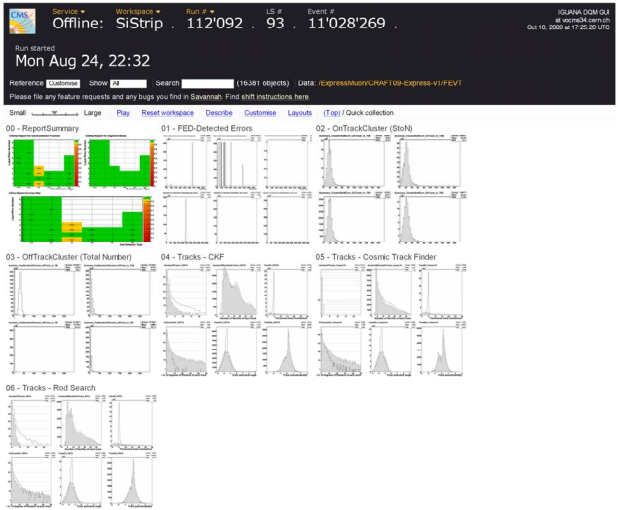

Fig. 2. Example view of the CMS DQM GUI, showing the SiStrip summary histograms, which are used by the CMS Tracker DQM offline shifter for a quick overview over the SiStrip data quality in a given run.

To accomplish to the two main DQM tasks described in section I., the CMS DQM runs independently in the online and offline environments:

### A. Online DQM

Since the production of histograms including reconstruction requires more time per event than available during data taking[1], the CMS DQM is run with a reduced rate in the online environment compared to the incoming event rate. This rate is defined individually for each sub-detector. However, the available computing resources allow for parallel running DQM processes, so that histograms can be filled up to the full detector granularity.

In the online environment, the task of detector monitoring and problem identification is emphasized.

### B. Offline DQM

The offline data processing performs reconstruction and DQM tasks for the full set of recorded data, so that the DQM obtains a more accurate picture by using all events to fill its histograms. However, since all DQM software modules run in one process here, memory limitations restrict the DQM to reduce the number of produced histograms.

In the offline environment, the task of data certification is emphasized, but also problems can be investigated in more detail compared to the online environment.

## IV. THE STRUCTURE OF THE CMS TRACKER DQM

The CMS Tracker DQM has to monitor data over a wide range of processing complexity, starting from integrity checks performed on raw (unreconstructed) data up to checks of the full track reconstruction. It is subdivided into three sub-systems: SiStrip and Pixel DQM for monitoring detector performances and low level reconstruction, and Tracking DQM, where Tracking is considered as the highest level of local reconstruction, taking into account data from both sub-detectors. Both sub-detector DQM systems use also reconstructed tracks to monitor the detector performances.

The hierarchical folder structure of the SiStrip and Pixel DQM output follows strictly the geometrical segmentation of the particular sub-detector. In addition, a parallel folder structure follows also the arrangement of the SiStrip data acquisition system (DAQ), since this is *not* identical to the geometrical structure. The mostly used type of histograms is the one-dimensional histogram (distribution), and it is filled for *all* levels of detector granularity. The aim to monitor data integrity and the different steps of local reconstruction efficiently up to the highest detector granularity leads to a total number of about 350,000 histograms for the sub-detectors. However, this level of detail is essential to perform the task of monitoring and problem identification sufficiently.

Summary histograms ease the navigation through this huge amount of histograms and provide sensitive information already at lower detector granularity levels. They are defined in such a way, that they display a particular property of histograms defined at higher detector granularity bin-by-bin per detector part, e.g. one bin for each detector module summarized for a SiStrip layer. Fig. 3 shows an example of a distribution at highest granularity and a corresponding summary at a reduced granularity level.

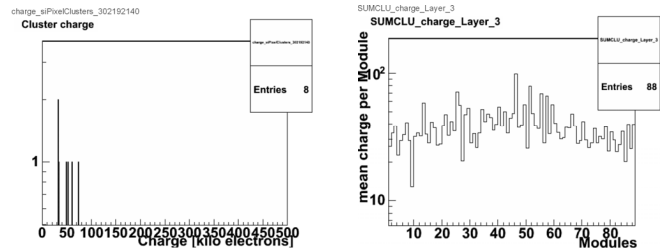

Fig. 3. An example of summary histograms. The left plot shows a distribution of cluster charges in a Pixel module (highest detector granularity). The right plot summarizes the mean values of all corresponding distributions for a Pixel layer. The left plot is represented by module 48 in the right plot.

Track reconstruction monitoring does not require a deep hierarchical structure of the histograms, since this reconstruction step uses the detector as a whole. However, histograms using reconstructed tracks are present for the sub-detectors up to the highest granularity level, e.g. the number of reconstructed clusters per module that have not been assigned to a reconstructed track. These histograms are particularly useful to identify noisy modules. The CMS Tracker DQM runs the whole sub-detector related reconstruction, namely the track reconstruction, *completely* already in the online environment to obtain these histograms.

Expert visualization tools have been developed for the CMS Tracker DQM in order to serve its special needs. These tools are briefly introduced here.

---

[1] CMS will record data at a rate of 100 Hz

## A. CMS Tracker DQM Expert GUI

The CMS Tracker Expert GUI is a visualization tool for histograms specialized for the deep hierarchical structure and large number of histograms in the Tracker case. It provides especially quick and easy navigation and complementary views like e.g. the "Alarm View", which points to histograms with failing quality tests. Fig. 4 shows an example view of the CMS Tracker Expert GUI for the Pixel. Details about the CMS Tracker DQM Expert GUI can be found in [3].

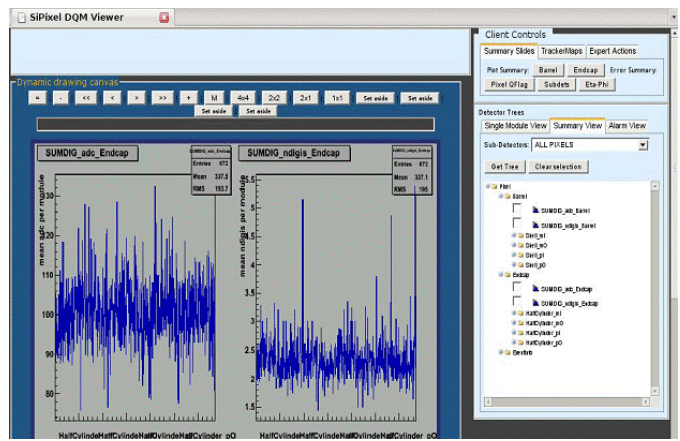

Fig. 4. Example view of the CMS Tracker Expert GUI. The navigation panel on the right gives the opportunity for extensive customization and to choose from several view modes.

## B. Tracker Maps

The so-called *Tracker Maps* are two-dimensional synoptic views, which display certain histogram properties, like e.g. the mean of a particular histogram, at highest detector granularity either in their geometrical or read-out (DAQ) context. This offers the possibility to identify patterns and helps in problem diagnosis. An example Tracker Map detail is shown in Fig. 5, which gives an example of a successful problem diagnosis with CMS Tracker DQM tools. Details about the Tracker Maps can be found in [3].

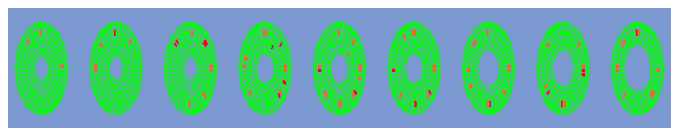

Fig. 5. Example Tracker Map detail showing the geometrical representation of quality test results of SiStrip modules in TEC. Green entries represent good modules, red entries represent modules with any failing quality test. In this case, wrong calibration constants led to increased noise in a number of modules. The fact, that some of these modules are in-line, caused the reconstruction of fake tracks along the LHC beam axis.

## C. Historic DQM

In order to monitor the evolution of detector properties over time, the Historic DQM (hDQM) tool has been developed by the CMS Tracker DQM group [4]. It basically consists of a database, a client and a web interface. Its working principle is shown in Fig. 6. Defined quantities of histograms are extracted from the DQM output after the offline data processing and are injected into the database for each run. From there, histograms showing the time evolution of these quantities (trend plots) can be produced. They can be used either to find outliers or to monitor long-term evolutions. An example plot is shown in Fig. 7. This tool has been adopted by the CMS DQM and is used by many CMS sub-detectors in the meanwhile.

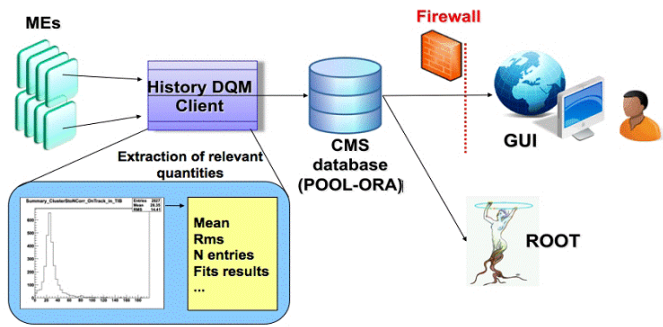

Fig. 6. Working principle of the hDQM. The DQM consumer extracts defined quantities from histograms and injects them into a database. From there, users have access and can produce plots of time evolutions.

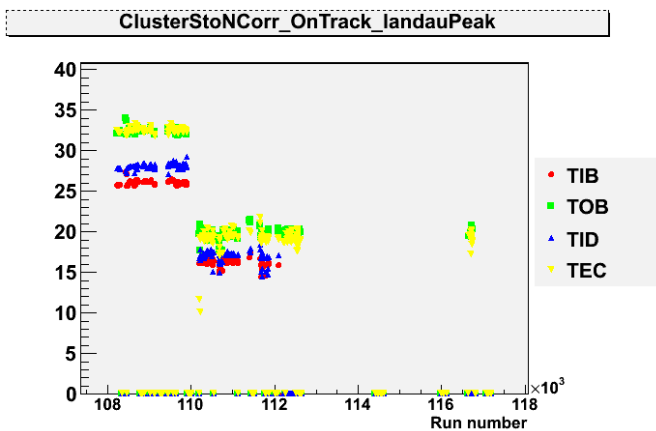

Fig. 7. Example hDQM plot showing the time evolution of the signal-to-noise ratio for the four SiStrip detector parts. A drop caused by a switch in the used read-out mode is easily seen.

## V. THE DATA CERTIFICATION WORK FLOW FOR THE CMS TRACKER

The CMS DQM uses generally two complementary approaches for data certification in parallel: (a) automatic certification with quality flags and (b) manual certification by shift personnel (*shifters*). This is done centrally and for each sub-system individually, where the particular task sharing differs slightly between the online and offline environment. Whereas the current emphasis during the commissioning phase of the CMS experiment is on the manual run-by-run certification, which relies on visual inspection of histograms, it will move to the automatic certification also with higher time resolution in the future with the manual one as control instance.

The CMS Tracker DQM uses all available tools for its data certification. After the DQM producers have produced the histograms, quality tests are applied to them and the quality flags are computed by the CMS Tracker DQM consumers. There are in total 13 flags for the Pixel, 20 for the SiStrip and four for the Tracking, where basic and combined flags are counted and also flags from other sources than DQM[2] are taken into account. The histograms are uploaded to the CMS DQM GUI, Tracker Maps are produced and in offline processing, the hDQM database is populated.

Generally, data is monitored by central CMS shifters as well as by expert shifters of the sub-systems. However, if a problem has been identified, only the expert shifters can trace the problem in detail and up to the highest detector granularity.

In the online environment, the data is assessed by the CMS DQM online shifter on the basis of quality flags and a small number of summary histograms. The results of this assessment are Boolean flags per sub-system telling if the current run can be used for physics analyses that use this sub-system, and they are written to the Run Registry database. The CMS Tracker online shifter also inspects histograms and in addition Tracker Maps, but her/his work emphasizes monitoring and problem identification for quick feedback. She/he does not determine quality flags, but any information that is useful for data certification is communicated to be used by the CMS Tracker offline shifters, as described later.

In the offline environment, the procedure for the CMS DQM shifter is similar, where her/his assessment supersedes that one of the online shifter, since it is based on fully reconstructed data. However, the CMS Tracker DQM offline shifters have the opportunity for a more detailed look into the DQM output, and they use Tracker Maps and hDQM in addition, the latter for the identification of outliers. This environment is also more sufficient for reconstruction monitoring and detailed problem tracing.

All shifters have also communicated findings of possible problems to CMS Tracker experts, who analyze them to find the cause. These expert analysis results are finally combined by the CMS Tracker offline shift leader with the outcome of all automatic and manual certification steps. The central CMS DQM is notified once a week of the final data certificates for the three CMS Tracker sub-systems, which takes all possible sources of information into account.

A sketch of the general data certification work flow for the CMS Tracker is shown in Fig. 8, summarizing contributions from the on- and offline environments.

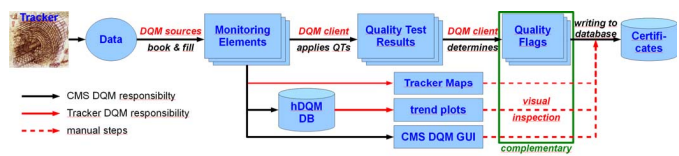

Fig. 8. General CMS Tracker DQM certification work flow. Red labels and arrows show the responsibilities of the CMS Tracker DQM compared to the CMS DQM: preparation of DQM producers and consumer code, creation of Tracker Maps, production of trend plots from the hDQM database and visual inspections performed by CMS Tracker shifters. The sketched structure combines both, online and offline contributions to the certification.

---

[2] Data Acquisition (DAQ) and Detector Control System (DCS)

## VI. CONCLUSION

The CMS Tracker DQM has been developed to monitor the detector and reconstruction performance efficiently and to certify the recorded data for the three CMS sub-systems Si-Strip, Pixel and Tracking. It is bound on the one hand to the specific detector designs and on the other hand to the structure of the central CMS DQM. Within these framework conditions, tools and work flows have been developed, which allow for quick analyses and solutions of problems and for reliable judgments on the usability of data. It has been an essential tool during the recent commissioning phase of the CMS experiment and is prepared to fulfill its task during physics data taking.

## ACKNOWLEDGMENT

I thank the whole CMS Tracker DQM group, Suchandra Dutta, Petra Merkel and Laura Borrello, for their support in preparing my contribution.

I also thank the Vrije Universiteit Brussel, especially Jorgen d'Hondt, for supporting my attendance to this conference.